# On the role of synaptic stochasticity in training low-precision neural networks

Carlo Baldassi,[1,2,3] Federica Gerace,[2,4] Hilbert J. Kappen,[5] Carlo Lucibello,[2,4] Luca Saglietti,[2,4] Enzo Tartaglione,[2,4] and Riccardo Zecchina[1,2,6]

[1] *Bocconi Institute for Data Science and Analytics, Bocconi University, Milano, Italy*
[2] *Italian Institute for Genomic Medicine, Torino, Italy*
[3] *Istituto Nazionale di Fisica Nucleare, Sezione di Torino, Italy*
[4] *Dept. of Applied Science and Technology, Politecnico di Torino, Torino, Italy*
[5] *Radboud University Nijmegen, Donders Institute for Brain, Cognition and Behaviour 6525 EZ Nijmegen, The Netherlands*
[6] *International Centre for Theoretical Physics, Trieste, Italy*

Stochasticity and limited precision of synaptic weights in neural network models are key aspects of both biological and hardware modeling of learning processes. Here we show that a neural network model with stochastic binary weights naturally gives prominence to exponentially rare dense regions of solutions with a number of desirable properties such as robustness and good generalization performance, while typical solutions are isolated and hard to find. Binary solutions of the standard perceptron problem are obtained from a simple gradient descent procedure on a set of real values parametrizing a probability distribution over the binary synapses. Both analytical and numerical results are presented. An algorithmic extension aimed at training discrete deep neural networks is also investigated.

Learning can be regarded as an optimization process over the connection weights of a neural network. In nature, synaptic weights are known to be plastic, low precision and unreliable, and it is an interesting issue to understand if this stochasticity can help learning or if it is an obstacle. The debate about this issue has a long history and is still unresolved (see [1] and references therein). Here, we provide quantitative evidence that the stochasticity associated with noisy low precision synapses can drive elementary supervised learning processes towards a particular type of solutions which, despite being rare, are robust to noise and generalize well — two crucial features for learning processes.

In recent years, multi-layer (*deep*) neural networks have gained prominence as powerful tools for tackling a large number of cognitive tasks [2]. In a $K$-class classification task, neural network architectures are typically trained as follows. For any input $x \in \mathcal{X}$ (the input space $\mathcal{X}$ typically being a tensor space) and for a given set of parameters $W$, called *synaptic weights*, the network defines a probability density function $P(y\,|\,x, W)$ over the $K$ possible outcomes. This is done through composition of affine transformations involving the synaptic weights $W$, element wise non-linear operators, and finally a softmax operator that turns the outcome of previous operations into a probability density function [3]. The weights $W$ are adjusted, in a supervised learning scenario, using a training set $\mathcal{D}$ of $M$ known input-output associations, $\mathcal{D} = \{(x^\mu, y^\mu)\}_{\mu=1}^{M}$. The learning problem is reframed into the problem of maximizing a log-likelihood $\tilde{\mathcal{L}}(W)$ over the synaptic weights $W$:

$$\max_W \ \tilde{\mathcal{L}}(W) := \sum_{(x,y)\in\mathcal{D}} \log P(y\,|\,x, W) \tag{1}$$

The maximization problem is approximately solved using variants of the Stochastic Gradient Descent (SGD) procedure over the loss function $-\tilde{\mathcal{L}}(W)$ [4]. In a Bayesian approach instead one is interested in computing the posterior distribution $P(W\,|\,\mathcal{D}) \propto P(\mathcal{D}\,|\,W)\,P(W)$, where $P(W)$ is some prior over the weights $W$. In deep networks, unfortunately, the exact computation of $P(W\,|\,\mathcal{D})$ is typically infeasible and various approximated approaches have been proposed [5–7].

Shallow neural network models, such as the perceptron model for binary classification, are amenable to analytic treatment while exposing a rich phenomenology. They have attracted great attention from the statistical physics community for many decades [8–16]. In the perceptron problem we have binary outputs $y \in \{-1, +1\}$, while inputs $x$ and weights $W$ are $N$-components vectors. Under some statistical assumptions on the training set $\mathcal{D}$ and for large $N$, single variable marginal probabilities $P(W_i \,|\, \mathcal{D})$ can be computed efficiently, using Belief Propagation [17–19]. The learning dynamics has also been analyzed, in particular in the online learning setting [11, 20]. In a slightly different perspective the perceptron problem is often framed as the task of minimizing the error-counting Hamiltonian

$$\min_W \;\; \mathcal{H}(W) \coloneqq \sum_{(x,y)\in\mathcal{D}} \Theta\left(-y\sum_{i=1}^{N} W_i\, x_i\right), \qquad (2)$$

where $\Theta(x)$ is the Heaviside step function, $\Theta(x) = 1$ if $x > 0$ and 0 otherwise. As a constraint satisfaction problem, it is said to be satisfiable (SAT) if zero energy (i.e. $\mathcal{H}(W) = 0$) configurations exists, unsatisfiable (UNSAT) otherwise. We call *solutions* such configurations. Statistical physics analysis, assuming random and uncorrelated $\mathcal{D}$, shows a sharp threshold at a certain $\alpha_c = M/N$, when $N$ grows large, separating a SAT phase from an UNSAT one. Moreover, restricting the synaptic space to binary values, $W_i = \pm 1$, leads to a more complex scenario: most solutions are essentially isolated and computationally hard to find [13, 21]. Some efficient algorithms do exist though [12, 22] and generally land in a region dense of solutions. This apparent inconsistency has been solved through a large deviation analysis which revealed the existence of sub-dominant and dense regions of solutions [14, 23]. This analysis introduced the concept of Local Entropy [14] which subsequently led to other algorithmic developments [24–26] (see also [27] for related analysis).

In the generalization perspective, solutions within a dense region may be loosely considered as representative of the entire region itself, and therefore act as better pointwise predictors than isolated solutions, since the optimal Bayesian predictor is obtained averaging all solutions [14].

Here, we propose a method to solve the binary perceptron problem (2) through a relaxation to a distributional space. We introduce a perceptron problem with stochastic discrete weights, and show how the learning process is naturally driven towards dense regions of solutions, even in the regime in which they are exponentially rare compared to the isolated ones. In perspective, the same approach can be extended to the general learning problem (1), as we will show.

Denote with $Q_\theta(W)$ a family of probability distributions over $W$ parametrized by a set of variables $\theta$. Consider the following problem:

$$\max_\theta \;\; \mathcal{L}(\theta) \coloneqq \sum_{(x,y)\in\mathcal{D}} \log \mathbb{E}_{W\sim Q_\theta} P(y \,|\, x, W) \qquad (3)$$

Here $\mathcal{L}(\theta)$ is the log-likelihood of a model where for each training example $(x, y) \in \mathcal{D}$ the synaptic weights are independently sampled according to $Q_\theta(W)$. Within this scheme two class predictors can be devised for any input $x$: $\hat{y}_1(x) = \mathrm{argmax}_y\, P(y \,|\, x, \hat{W})$, where $\hat{W} = \mathrm{argmax}_W\, Q_\theta(W)$, and $\hat{y}_2(x) = \mathrm{argmax}_y \int \mathrm{d}W\, P(y \,|\, x, W)\, Q_\theta(W)$. In this paper we will analyze the quality of the training error given by the first predictor. Generally, dealing with Problem (3) is more difficult than dealing with Problem (1), since it retains some of the difficulties of the computation of $P(W \,|\, \mathcal{D})$. Also notice that for any maximizer $W^\star$ of Problem (1) we have that $\delta(W - W^\star)$ is a maximizer of Problem (3) provided that it belongs to the parametric family, as can be shown using Jensen's inequality. Problem (3) is a "distributional" relaxation of Problem (1).

Optimizing $\mathcal{L}(\theta)$ instead of $\tilde{\mathcal{L}}(W)$ may seem an unnecessary complication. In this paper we argue that there are two reasons for dealing with this kind of task. First, when the configuration space of each synapse is restricted to discrete values, the network cannot be trained with SGD procedures. The problem, while being very important for computational efficiency and memory gains, has been tackled only very recently [5, 28]. Since variables $\theta$ typically lie in a continuous manifold instead, standard continuous optimization tools can be applied to $\mathcal{L}(\theta)$. Also, the learning dynamics on $\mathcal{L}(\theta)$ enjoys some additional properties when compared to the dynamics on $\tilde{\mathcal{L}}(W)$. In the latter case additional regularizers, such as dropout and $L_2$ norm, are commonly used to improve generalization properties [4]. The SGD in the $\theta$-space instead already incorporates the kind of natural regularization intrinsic in the Bayesian approach and the robustness associated to high local entropy [14]. Here we make a case for these arguments by a numerical and analytical study of the proposed approach for the binary perceptron. We also present promising preliminary numerical results on deeper networks.

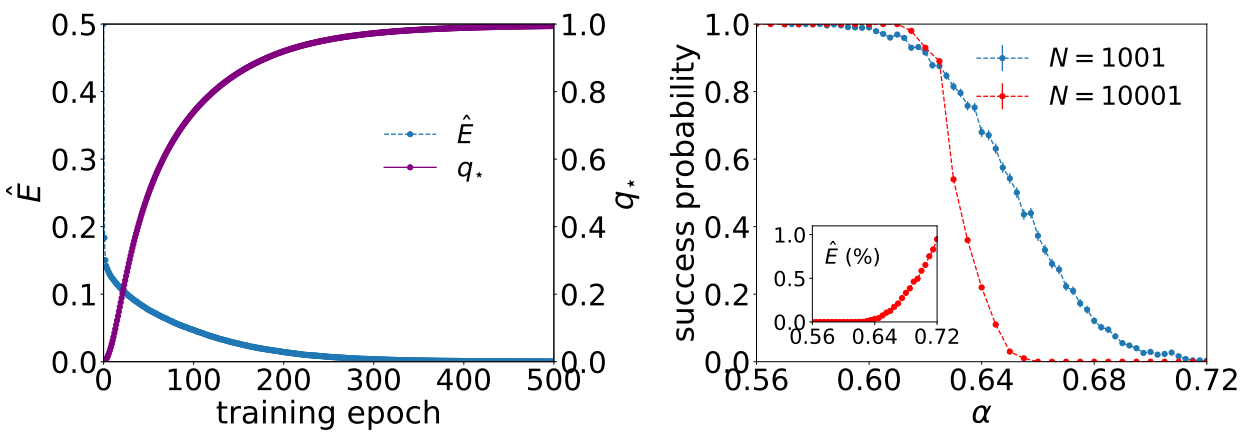

Figure 1. (*Left*) The training error and the squared norm against the number of training epochs, for $\alpha = 0.55$ and $N = 10001$, averaged over 100 samples. (*Right*) Success probability in the classification task as a function of the load $\alpha$ for networks of size $N = 1001, 10001$ averaging 1000 and 100 samples respectively. In the inset we show the average training error at the end of GD as a function of $\alpha$.

*Learning for the Stochastic Perceptron.* Following the above discussion, we now introduce our binary stochastic perceptron model. For each input $x$ presented, $N$ synaptic weights $W = (W_1, \ldots, W_N)$, $W_i \in \{-1, +1\}$, are randomly extracted according to the distribution

$$Q_m(W) = \prod_{i=1}^{N} \left[ \frac{1+m_i}{2} \delta_{W_i,+1} + \frac{1-m_i}{2} \delta_{W_i,-1} \right] \quad (4)$$

where $\delta_{a,b}$ is the Kronecker delta symbol. We will refer to the set $m = (m_i)_i$, where $m_i \in [-1, 1]$ $\forall i$, as the magnetizations or the control parameters. We choose the probability $P(y \,|\, x, W)$ on the class $y \in \{-1, +1\}$ for a given input $x$ as follows:

$$P(y \,|\, x, W) = \Theta\left( y \sum_{i=1}^{N} W_i \, x_i \right). \quad (5)$$

While other possibilities for $P(y \,|\, x, W)$ could be considered, this particular choice is directly related to the form of the Hamiltonian in Problem (2), which we ultimately aim to solve. Given a training set $\mathcal{D} = \{(x^\mu, y^\mu)\}_{\mu=1}^{M}$, we can then compute the log-likelihood function of Eq. (3), with the additional assumption that $N$ is large and the central limit theorem applies. It reads

$$\mathcal{L}(m) = \sum_{(x,y)\in\mathcal{D}} \log H\left( -\frac{y \sum_i m_i \, x_i}{\sqrt{\sum_i \left(1 - m_i^2\right) x_i^2}} \right), \quad (6)$$

where $H(x) := \int_x^\infty dz \, e^{-z^2/2}/\sqrt{2\pi}$. Minimizing $-\mathcal{L}(m)$ instead of finding the solutions of Problem (2) allows us to use the simplest method for approximately solving continuous optimization problems, the Gradient Descent (GD) algorithm:

$$m_i^{t+1} \leftarrow \text{clip}\left( m_i^t + \eta \, \partial_{m_i} \mathcal{L}\left(m^t\right) \right). \quad (7)$$

We could have adopted the more efficient SGD approach, however in our case simple GD is already effective. In the last equation $\eta$ is a suitable learning rate and $\text{clip}(x) := \max(-1, \min(1, x))$, applied element-wise. Parameters are randomly initialized to small values, $m_i^0 \sim \mathcal{N}\left(0, N^{-1}\right)$. At any epoch $t$ in the GD dynamics a binarized configuration $\hat{W}_i^t = \text{sign}\left(m_i^t\right)$ can be used to compute the training error $\hat{E}^t = \frac{1}{M} \mathcal{H}\left(\hat{W}^t\right)$. We consider a training set $\mathcal{D}$ where each input component $x_i^\mu$ is sampled uniformly and independently in $\{-1, 1\}$ (with this choice we can set $y^\mu = 1 \, \forall \mu$ without loss of generality). The evolution of the network during GD is shown in Fig. 1. The training error goes progressively to zero while the mean squared norm of the control variables $q_\star^t = \frac{1}{N} \sum_i \left(m_i^t\right)^2$ approaches one. Therefore the distribution $Q_m$ concentrates around a single configuration as the training is progressing. This natural flow is similar to the annealing of the coupling parameter manually performed in local entropy inspired algorithms [25, 26]. We also show in Fig. 1 the probability over the realizations of $\mathcal{D}$ of finding a solution of the binary problem as function of the load $\alpha = M/N$. The algorithmic capacity of GD is approximately $\alpha_{GD} \approx 0.63$. This value has to be compared to the theoretical capacity $\alpha_c \approx 0.83$, above which there are almost surely no solutions [9], and state-of-the-art algorithms based on message passing heuristics for which we have a range of capacities $\alpha_{MP} \in [0.6, 0.74]$ [12, 22, 29]. Therefore GD reaches loads only slightly worse than those reached by much more fine tuned algorithms, a surprising results for such a simple procedure. Also, for $\alpha$ slightly above $\alpha_{GD}$ the training error remains comparably low, as shown in Fig. 1. In our experiments most variants of the GD procedure of Eq. (7) performed just as well: e.g. SGD ors GD computed on the fields $h_i^t = \tanh^{-1}(m_i^t)$ rather than the magnetizations[30]. Other updates rules for the control parameters can be derived as multiple pass of on-line Bayesian learning [31, 32]. Variations of rule (7) towards biological plausibility are discussed in the SM.

*Deep Networks.* We applied our framework to deep neural networks with binary stochastic weights and sign activation functions. Using an uncorrelated neuron approximation, as in Ref. [6], we trained the network using the standard SGD algorithm with backpropagation. We give the details in the SM. On the MNIST benchmark problem [33], using a network with three hidden layers we achieved $\sim 1.7\%$ test error, a very good result for a network with binary weights and activations and with no convolutional layers [34]. No other existing approach to the binary perceptron problem has been extended yet to deeper settings.

*Statistical mechanics Analysis.* We now proceed with the analytical investigation of the equilibrium properties of the stochastic perceptron, which partly motivates the good performance of the GD dynamics. The starting point of the analysis is the partition function

$$Z = \int_{\Omega} \prod_i \mathrm{d}m_i \; \delta\left(\sum_i m_i^2 - q_\star N\right) e^{\beta \mathcal{L}(m)} \qquad (8)$$

where $\Omega = [-1,1]^N$, $\beta$ is an inverse temperature, and we constrained the squared norm to $q_\star N$ in order to mimic the natural flow of $q_\star^t$ in the training process. The dependence on the training set $\mathcal{D}$ is implicit in last equation. We shall denote with $\mathbb{E}_{\mathcal{D}}$ the average over the training sets with i.i.d. input and output components uniform in $\{-1,1\}$. We investigate the average properties of the system for large $N$ and fixed load $\alpha = M/N$ using the replica method in the Replica Symmetric (RS) ansatz [35]. Unfortunately the RS solution becomes locally unstable for very large $\beta$. Therefore, instead of taking the infinite $\beta$ limit to maximize the likelihood we will present the results obtained for $\beta$ large but still in the RS region. The details of the free energy calculation and of the stability check can be found in the SM.

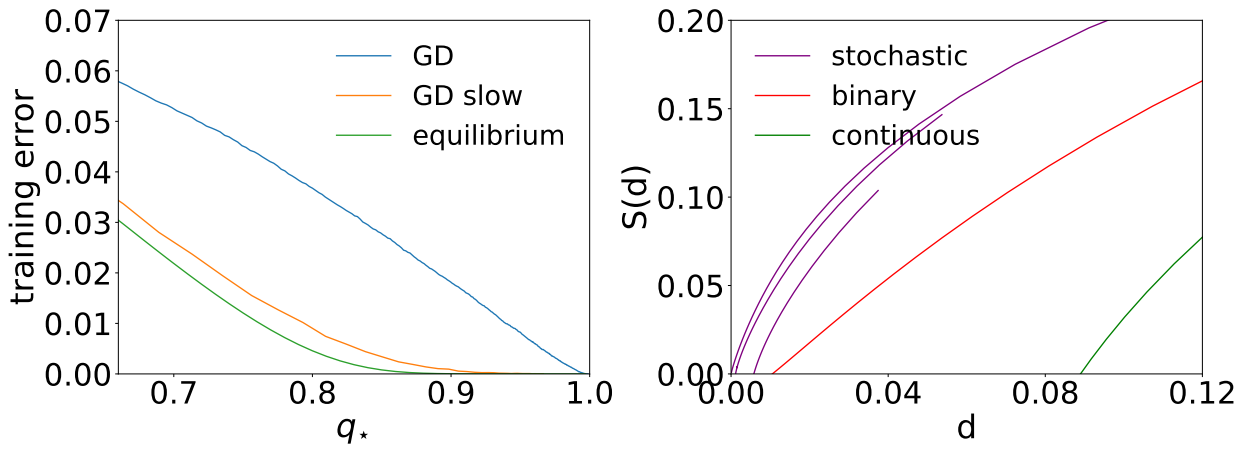

Figure 2. (*Left*) Energy of the Binarized Configuration versus the control variable $q_\star$. We show the equilibrium prediction of Eq. (9), and numerical results from the GD algorithm and a GD algorithm variant where after each update we rescale the norm of $m$ to $q_\star$ until convergence before moving to the next value of $q_\star$ according to a certain schedule. The results are averaged over 20 random realizations of the training set with $N = 10001$. (*Right*) Entropy of binary solutions at fixed distance $d$ from BCs of the spherical, binary and stochastic perceptron ($q_\star = 0.7$, 0.8 and 0.9 from bottom to top) at thermodynamic equilibrium. In both figures $\alpha = 0.55$, also $\beta = 20$ for the stochastic perceptron and $\beta = \infty$ for the spherical and binary ones.

*Energy of the Binarized Configuration.* We now analyze some properties of the mode of the distribution $Q_m(W)$, namely $\hat{W}_i = \mathrm{sign}(m_i)$, that we call Binarized Configuration (BC). The average training error per pattern is:

$$E = \lim_{N\to\infty} \frac{1}{\alpha N} \mathbb{E}_{\mathcal{D}} \left[ \sum_{(x,y)\in\mathcal{D}} \left\langle \Theta\left(-y \sum_i \mathrm{sign}(m_i)\, x_i\right) \right\rangle \right] \qquad (9)$$

where $\langle \bullet \rangle$ is the thermal average over $m$ according to the partition function (8), which implicitly depends on $\mathcal{D}$, $q_\star$ and $\beta$. The last equation can be computed analytically within the replica framework (see SM). In Fig. 2 (*Left*) we show that for large $\beta$ the BC becomes a solution of the problem when $q_\star$ approaches one. This is compared to the values of the training error obtained from GD dynamics at corresponding values of $q_\star$, and a modified GD dynamics where we let the system equilibrate at fixed $q_\star$. The latter case, although we are at finite $N$ and we are considering a dynamical process that could suffer the presence of local minima, is in rea-sonable agreement with the equilibrium result of Eq. (9).

*Geometrical structure of the solution space.* Most solutions of the binary perceptron problem are isolated [13], while a subdominant but still exponentially large number belongs to a dense connected region [14]. Solutions in the dense region are the only ones that are algorithmically accessible. Here we show that BCs of the stochastic binary perceptron typically belong to the dense region, provided $q_\star$ is high enough. To prove this we count the number of solutions at a fixed Hamming distance $d$ from typical BC (this corresponds to fixing an overlap $p = 1 - 2d$). Following the approach of Franz and Parisi [36] we introduce the constrained partition function

$$\mathcal{Z}(d, m) = \sum_{W} \prod_{(x,y)\in\mathcal{D}} \Theta\left(y \sum_i W_i x_i\right) \times \delta\left(N\left(1-2d\right) - \sum_i \mathrm{sign}\left(m_i\right) W_i\right), \quad (10)$$

where the sum is over the $\{-1, +1\}^N$ binary configurations. The Franz-Parisi entropy $\mathcal{S}\left(d\right)$ is then given by

$$\mathcal{S}(d) = \lim_{N\to\infty} \frac{1}{N} \mathbb{E}_{\mathcal{D}} \left\langle \log \mathcal{Z}\left(d, m\right)\right\rangle. \quad (11)$$

We show how to compute $\mathcal{S}\left(d\right)$ in the SM. In Fig. 2 (*Right*) we compare $\mathcal{S}\left(d\right)$ for the stochastic perceptron with the analogous entropies obtained substituting the expectation $\langle \bullet \rangle$ over $m$ in Eq. (11) with a uniform sampling from the solution space of the spherical (the model of Ref. [8]) and the binary (as in Ref. [13]) perceptron. The distance gap between the BC and the nearest binary solutions (i.e., the value of the distance after which $\mathcal{S}(d)$ becomes positive) vanishes as $q_\star$ is increased: in this regime the BC belongs to the dense cluster and we have an exponential number of solutions at any distance $d > 0$. Typical binary solutions and binarized solutions of the continuous perceptron are isolated instead (finite gap, corresponding to $S(d) = 0$ at small distances). In the SM we provide additional numerical results on the properties of the energetic landscape in the neighborhood of different types of solutions, showing that solutions in flatter basins achieve better generalization than those in sharp ones.

*Conclusions.* Our analysis shows that stochasticity in the synaptic connections may play a fundamental role in learning processes, by effectively reweighting the error loss function, enhancing dense, robust regions, suppressing narrow local minima and improving generalization.

The simple perceptron model allowed us to derive analytical results as well as to perform numerical tests. Moreover, as we show in the SM, when considering discretized priors, there exist a connection with the dropout procedure which is popular in modern deep learning practice. However, the most promising immediate application is in the deep learning scenario, where this framework can be extended adapting the tools developed in Refs. [6, 7], and where we already achieved state-of-the-art results in our preliminary investigations.

Hopefully, the general mechanism shown here can also help to shed some light on biological learning processes, where the role of low precision and stochasticity is still an open question. Finally, we note that this procedure is not limited to neural network models; for instance, application to constraint satisfaction problems is straightforward.

CB, HJB and RZ acknowledge ONR Grant N00014-17-1-2569.

# On the role of synaptic stochasticity in training low-precision neural networks
# Supplementary Material

Carlo Baldassi,[1,2,3] Federica Gerace,[2,4] Hilbert J. Kappen,[5] Carlo Lucibello,[2,4] Luca Saglietti,[2,4] Enzo Tartaglione,[2,4] and Riccardo Zecchina[1,2,6]

[1] *Bocconi Institute for Data Science and Analytics, Bocconi University, Milano, Italy*
[2] *Italian Institute for Genomic Medicine, Torino, Italy*
[3] *Istituto Nazionale di Fisica Nucleare, Sezione di Torino, Italy*
[4] *Dept. of Applied Science and Technology, Politecnico di Torino, Torino, Italy*
[5] *Radboud University Nijmegen, Donders Institute for Brain, Cognition and Behaviour 6525 EZ Nijmegen, The Netherlands*
[6] *International Centre for Theoretical Physics, Trieste, Italy*

## CONTENTS

## I. REPLICA SYMMETRIC SOLUTION AND STABILITY ANALYSIS

In this Section we show how to compute the average the free entropy of the stochastic perceptron model discussed in the main text, using the replica trick and under Replica Symmetry (RS) assumptions. The limit of validity of the RS ansatz will also be discussed. The statistical physics model of a perceptron with $N$ stochastic binary synapses is defined by the partition function:

$$Z_N = \int_\Omega \prod_{i=1}^N \mathrm{d}m_i \; \delta\left(\sum_{i=1}^N m_i^2 - q_\star N\right) e^{\beta \mathcal{L}(m)}. \tag{1}$$

The partition function depends implicitly on the inverse temperature $\beta$, a training set $\mathcal{D} = \{y^\mu, x^\mu\}_{\mu=1}^M$, $M = \alpha N$ for some $\alpha > 0$, $y^\mu \in \{-1, 1\}$, $x^\mu \in \{-1, 1\}^N$, and a norm parameter $q_\star$. The integration is in the box $\Omega = [-1, 1]^N$. The log-likelihood is given, for large $N$, by

$$\mathcal{L}(m) = \sum_{\mu=1}^{M} \log H\left(-\frac{y^\mu \sum_{i=1}^N m_i\, x_i^\mu}{\sqrt{N(1-q_\star)}}\right), \tag{2}$$

with $H(x) = \frac{1}{2}\operatorname{erfc}\left(\frac{x}{\sqrt{2}}\right)$. As usual in statistical physics, we are interested in the large $N$ limit of the system, an assumption which has already been exploited in expressing $\mathcal{L}(m)$ as in Eq. (2), as already stated. Also, the average $\mathbb{E}_\mathcal{D}$ over random instances of the training set is considered: $x_i^\mu$ are uniformly and independently distributed over $\{-1, 1\}$ and without loss of generality we set $y^\mu = 1\ \forall \mu$. Notice that although $\mathcal{L}(m)$ is concave on the box $\Omega$, the norm constraint decomposes $\Omega$ into disjoint domains, therefore the maximization problem (i.e. the large $\beta$ limit) is non-trivial.

We define the average asymptotic free entropy as

$$\phi = \lim_{N\to\infty} \frac{1}{N} \mathbb{E}_\mathcal{D} \log Z_N \tag{3}$$

where the limit is taken at fixed $\alpha$. In order to compute $\phi$ we shall resort to the standard machinery of the replica method [13, 20]. The replicated partition function of the model is given by

$$\mathbb{E}_\mathcal{D} Z_N^n = \mathbb{E}_\mathcal{D} \int_{\Omega^{\otimes n}} \prod_{a=1}^n \prod_{i=1}^N \mathrm{d}m_i^a \prod_{a=1}^n \delta\left(\sum_{i=1}^N (m_i^a)^2 - q_\star N\right) \prod_{\mu=1}^M \prod_{a=1}^n H^\beta\left(-\frac{\sum_{i=1}^N x_i^\mu m_i^a}{\sqrt{N(1-q_\star)}}\right). \tag{4}$$

After some manipulations, at the leading order in $N$, we obtain

$$\mathbb{E}_\mathcal{D} Z_N^n \sim \int \prod_a \frac{\mathrm{d}\hat{q}_{aa}}{2\pi} \prod_{a<b} \frac{\mathrm{d}\hat{q}_{ab}\mathrm{d}q_{ab}}{2\pi} e^{Nn\phi[\hat{q},q]}, \tag{5}$$

where the replicated free entropy is given by

$$\phi[\hat{q}, q] = -\frac{1}{2n}\sum_{a,b} \hat{q}_{ab}\, q_{ab} + G_S[\hat{q}] + \alpha G_E[q], \tag{6}$$

where we defined $q_{aa} \equiv q_\star$ for convenience. In last equation we also defined

$$G_S[\hat{q}] = \frac{1}{n}\log \int_{[-1,1]^n} \prod_a \mathrm{d}m_a\; e^{\frac{1}{2}\sum_{ab}\hat{q}_{ab} m_a m_b}, \tag{7}$$

$$G_E[q] = \frac{1}{n}\log \int \prod_a \frac{\mathrm{d}\hat{u}_a \mathrm{d}u_a}{2\pi}\; e^{-\frac{1}{2}\sum_{ab} q_{ab}\hat{u}_a\hat{u}_b + i\hat{u}_a u_a} \prod_a H^\beta\left(-\frac{u_a}{\sqrt{1-q_\star}}\right). \tag{8}$$

Saddle point evaluation of the replicated partition function yields the following identities:

$$\hat{q}_{ab} = -\alpha \left\langle\left\langle \hat{u}_a \hat{u}_b \right\rangle\right\rangle_E \qquad a > b, \tag{9}$$

$$q_{ab} = \left\langle\left\langle m_a m_b \right\rangle\right\rangle_S \qquad a > b, \tag{10}$$

$$q_{aa} \equiv q_\star = \left\langle\left\langle m_a^2 \right\rangle\right\rangle_S . \tag{11}$$

Here we denoted with $\langle\langle\bullet\rangle\rangle_S$ and $\langle\langle\bullet\rangle\rangle_E$ the expectations taken according to the single-body partition function in the logarithms of Eq. (7) and Eq. (8) respectively. Notice that last equation is an implicit equation for $\hat{q}_{aa}$.

We perform the saddle point evaluation and analytic continuation of $n \downarrow 0$ within the Replica Symmetric (RS) ansatz. Therefore we have $q_{ab} = q_0$, $\hat{q}_{ab} = \hat{q}_0$ for $a \neq b$ and $\hat{q}_{aa} = \hat{q}_1$ $\forall a$. The RS prediction for the average free entropy is then given by

$$\phi_{RS} = \underset{q_0,\hat{q}_0,\hat{q}_1}{\text{extr}} \quad \frac{1}{2}\left(q_0\hat{q}_0 - q_\star\hat{q}_1\right) + G_S\left(\hat{q}_0,\hat{q}_1\right) + \alpha G_E\left(q_0, q_\star\right) \tag{12}$$

where

$$G_S\left(\hat{q}_0,\hat{q}_1\right) = \int \mathcal{D}z \log \int_{-1}^{1} \mathrm{d}m \; e^{\frac{1}{2}(\hat{q}_1-\hat{q}_0)m^2 + \sqrt{\hat{q}_0} z m}, \tag{13}$$

$$G_E\left(q_0\right) = \int \mathcal{D}z \log \int \mathcal{D}u \; H^\beta\left(-\frac{\sqrt{q_0}z + \sqrt{q_\star - q_0}u}{\sqrt{1-q_\star}}\right). \tag{14}$$

In last equation we used the notation $\int \mathcal{D}z = \int \frac{\mathrm{d}z}{\sqrt{2\pi}} e^{-\frac{z^2}{2}}$. Saddle point conditions yield the set of equations

$$q_0 = -2\frac{\partial G_S}{\partial \hat{q}_0}; \qquad \hat{q}_0 = -2\alpha\frac{\partial G_E}{\partial q_0}; \qquad 0 = 2\frac{\partial G_S}{\partial \hat{q}_1} - q_\star, \tag{15}$$

that we solve iteratively. Last equation is an implicit equation for $\hat{q}_1$.

This derivation is the starting point for the more complicated calculations of the energy of the Binarized Configurations (BCs) in Section II and of the Franz-Parisi entropy in Section III.

While we conjecture $\phi_{RS}$ to be exact at low $\beta$, in the region of high $\beta$ that we need to explore in order to maximize the log-likelihood $\mathcal{L}(m)$ it may be necessary to use a replica symmetry breaking formalism. A necessary, but not sufficient, condition for the validity of the RS formalism is the local stability condition for the free energy functional of Eq. (6) at the RS stationary point. The stability criterion involving the eigenvalues of the Hessian can be rephrased, with a slight adaption of the derivation of Ref. [14], as

$$\alpha \gamma_E \gamma_S < 1. \tag{16}$$

Here $\gamma_E$ and $\gamma_S$ are the relevant eigenvalues of the Hessians of $G_E[q]$ and $G_S[\hat{q}]$ respectively and for small $n$. They are given by

$$\gamma_E = \int \mathcal{D}z \left[\overline{\hat{u}^2}(z) - \left(\overline{\hat{u}}\left(z\right)\right)^2\right]^2, \tag{17}$$

$$\gamma_S = \int \mathcal{D}z \left[\overline{m^2}\left(z\right) - \left(\overline{m}\left(z\right)\right)^2\right]^2. \tag{18}$$

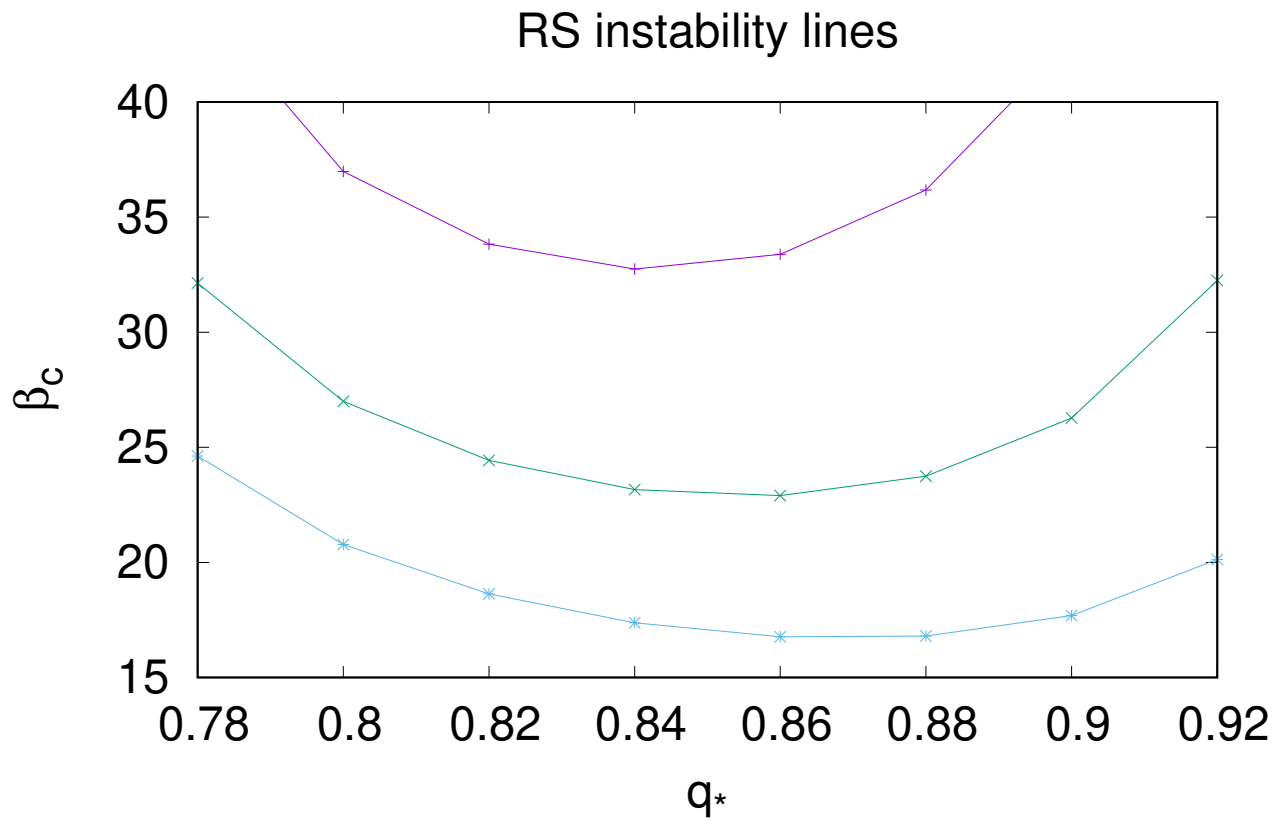

Figure 1. Critical value $\beta_c$ for the stability of the RS solution for different loads $\alpha = 0.5, 0.55, 0.6$ (from top to bottom) as a function of $q_\star$. Above $\beta_c$ the RS solution is unstable and a replica symmetry breaking ansatz should be considered to obtain the correct solution.

Expectations in lasts equations are defined by

$$\overline{\hat{u}^k}\left(z\right) \equiv \frac{\int \frac{\mathrm{d}\hat{u}\mathrm{d}u}{2\pi}\, \hat{u}^k\, e^{-\frac{1}{2}(q_\star - q_0)\hat{u}^2 + i\hat{u}u + i\hat{u}\sqrt{q_0}z} H^\beta\left(u\right)}{\int \frac{\mathrm{d}\hat{u}\mathrm{d}u}{2\pi}\, e^{-\frac{1}{2}(q_\star - q_0)\hat{u}^2 + i\hat{u}u + i\hat{u}\sqrt{q_0}z} H^\beta\left(u\right)} \tag{19}$$

$$\overline{m^k}\left(z\right) \equiv \frac{\int_{-1}^{1} \mathrm{d}m\; m^k\, e^{\frac{1}{2}(\hat{q}_1 - \hat{q}_0)m^2 + \sqrt{\hat{q}_0}zm}}{\int_{-1}^{1} \mathrm{d}m\; e^{\frac{1}{2}(\hat{q}_1 - \hat{q}_0)m^2 + \sqrt{\hat{q}_0}zm}} \tag{20}$$

In Fig. 1 we show the stability line $\beta_c\left(q_\star\right)$, defined by the condition $\alpha\gamma_E\gamma_S = 1$, for different values of $\alpha$. Due to numerical problem arising in computing integrals at high $\beta$, we explore a small $q_\star$ window. We note that $\beta_c\left(q_\star\right)$ stays finite in the range of parameters we explored and that the $\beta \uparrow \infty$ limit of the RS solution cannot be taken carelessly. Nonetheless the $\beta_c\left(q_\star\right)$ is generally quite high, although decreasing with $\alpha$. In the main text, where we presented the results for $\alpha = 0.55$, we set the inverse temperature to $\beta = 20$, where the RS results are supposedly correct and quantitatively close to the $\beta = +\infty$ limit.

## II. ENERGY OF THE BINARIZED CONFIGURATION

We now show how to compute the average energy $E$ (also called training error) associated to a typical Binarized Configuration (BC). In the thermodynamic limit it is written as

$$E = \lim_{N\to\infty} \mathbb{E}_D\left[\left\langle \Theta\left(-y^1 \sum_i \mathrm{sign}\left(m_i\right) x_i^1\right)\right\rangle\right], \tag{21}$$

where the average $\langle \bullet \rangle$ is over $m$ sampled according to the partition function (1) and $\Theta(x)$ is the Heaviside step function. Along the lines of previous Section, we resort to the replica approach, although here the replica of index 1 is distinguished from the others:

$$E = \lim_{N\to\infty} \lim_{n\to 0} \mathbb{E}_D \left[ \int_{\Omega^{\otimes n}} \prod_{a,i} \mathrm{d}m_i^a \prod_{a=1}^{n} \delta\left( \sum_{i=1}^{N} (m_i^a)^2 - q_\star N \right) \Theta\left( -\sum_i \mathrm{sign}\left(m_i^1\right) x_i^1 \right) e^{\beta \sum_a \mathcal{L}(m^a)} \right]. \tag{22}$$

In addition to the order parameters $q_{ab} = \frac{1}{N}\sum_i m_i^a m_i^b$ of Section I and the conjugated Lagrangian multipliers $\hat{q}_{ab}$, we also have to introduce the overlaps $p_a = \frac{1}{N}\sum_i \mathrm{sign}\left(m_i^1\right) m_i^a$ and the corresponding multipliers $\hat{p}_a$. We obtain the following expression for the mean energy $E$:

$$E = \lim_{N\to\infty} \lim_{n\to 0} \int \prod_{a<b} \mathrm{d}q_{ab} \prod_{a\leq b} \frac{\mathrm{d}\hat{q}_{ab}}{2\pi} \prod_a \frac{\mathrm{d}p_a \mathrm{d}\hat{p}_a}{2\pi} \; e^{Nn\tilde{\phi}[q,\hat{q},p,\hat{p}]} \, \tilde{E}\left[q,p\right] \tag{23}$$

The free entropy functional $\phi\left[q,\hat{q},p,\hat{p}\right]$ in this case reads

$$\tilde{\phi}\left[q,\hat{q},p,\hat{p}\right] = -\frac{1}{2n}\sum_{a,b} \hat{q}_{ab} q_{ab} - \frac{1}{n}\sum_a \hat{p}_a p_a + G_S\left[\hat{q},\hat{p}\right] + \alpha G_E\left[q\right] \tag{24}$$

$$G_S\left[\hat{q},\hat{p}\right] = \frac{1}{n}\log \int_{[-1,1]^n} \prod_a \mathrm{d}m^a \exp\left( \frac{1}{2}\sum_{a,b} \hat{q}_{ab} m^a m^b + \mathrm{sign}\left(m^1\right) \sum_a \hat{p}_a m^a \right) \tag{25}$$

$$G_E\left[q\right] = \frac{1}{n}\log \int \prod_a \frac{\mathrm{d}u_a \mathrm{d}\hat{u}_a}{2\pi} \prod_a H^\beta\left( -\frac{u_a}{\sqrt{1-q_\star}} \right) \exp\left( i\sum_a u_a \hat{u}_a - \frac{1}{2}\sum_{a,b} q_{ab}\hat{u}_a \hat{u}_b \right). \tag{26}$$

and the other term appearing in the integrand is given by

$$\tilde{E}\left[q,p\right] = \int \prod_a \frac{du_a d\hat{u}_a}{2\pi} \int \frac{d\tilde{u} d\hat{\tilde{u}}}{2\pi} \prod_a H^\beta\left( -\frac{u_a}{\sqrt{1-q_\star}} \right) \Theta\left(-\tilde{u}\right) \; e^{i\sum_a u_a \hat{u}_a - \frac{1}{2}\hat{\tilde{u}}^2 + i\tilde{u}\hat{\tilde{u}} - \frac{1}{2}\sum_{a,b} q_{ab}\hat{u}_a\hat{u}_b - \hat{\tilde{u}}\sum_a p_a \hat{u}_a}. \tag{27}$$

Saddle point evaluation of $\tilde{\phi}$ with respect to $p_a$ readily gives $\hat{p}_a = 0$. On this submanifold, $\tilde{\phi}$ reduces to the functional $\phi$ of previous Section, the matrix $q_{ab}$ and $\hat{q}_{ab}$ can be evaluated at saddle point in terms of $q_0, \hat{q}_1, \hat{q}_0$ within the RS ansatz and analytic continuation to $n = 0$ is finally obtained. Saddle point conditions with respect to $\hat{p}_a$ instead, i.e. $\partial\tilde{\phi}/\partial\hat{p}_a = 0$, fix the parameters $p_a \equiv \tilde{p}\ \forall a > 1$ and $p_1 \equiv p$ (here there is a little abuse of notation, the scalar value $p$ has not to be confused to the $n$-dimensional vector of previous equations). In conclusion, and in the small $n$ limit, after solving Eqs. (15) for $q_0, \hat{q}_1, \hat{q}_0$ we compute the saddle point values of $p$ and $\tilde{p}$ by

$$p = \int \mathcal{D}z \frac{\int_{-1}^{1} \mathrm{d}m \; \mathrm{sign}\left(m\right) m \, e^{\frac{1}{2}(\hat{q}_* - \hat{q}_0)m^2 + \sqrt{\hat{q}_0} z m}}{\int_{-1}^{1} \mathrm{d}m \, e^{\frac{1}{2}(\hat{q}_* - \hat{q}_0)m^2 + \sqrt{\hat{q}_0} z m}}, \tag{28}$$

$$\tilde{p} = \int \mathcal{D}z \frac{\left( \int_{-1}^{1} \mathrm{d}m \; e^{\frac{1}{2}(\hat{q}_* - \hat{q}_0)m^2 + \sqrt{\hat{q}_0} z_0 m} \, \mathrm{sign}\left(m\right) \right) \left( \int_{-1}^{1} \mathrm{d}m \; m \, e^{\frac{1}{2}(\hat{q}_* - \hat{q}_0)m^2 + \sqrt{\hat{q}_0} z m} \right)}{\left[ \int_{-1}^{1} \mathrm{d}m \, e^{\frac{1}{2}(\hat{q}_* - \hat{q}_0)m^2 + \sqrt{\hat{q}_0} z m} \right]^2}. \tag{29}$$

The value of $E$ is then simply given by $\tilde{E}$ evaluated on the saddle point. After some manipulation of the integrals appearing in Eq. (27) we finally arrive to

$$E = \int \mathcal{D}z \frac{\int \mathcal{D}u \; H\left(\frac{\frac{p-\tilde{p}}{\sqrt{q_\star - q_0}}u + \frac{\tilde{p}}{\sqrt{q_0}}z}{\sqrt{1-\frac{\tilde{p}^2}{q_0} - \frac{(p-\tilde{p})^2}{q_\star - q_0}}}\right) H^\beta\left(-\frac{\sqrt{q_\star - q_0}u + \sqrt{q_0}z}{\sqrt{1-q_\star}}\right)}{\int \mathcal{D}u \; H^\beta\left(-\frac{\sqrt{q_\star - q_0}u + \sqrt{q_0}z}{\sqrt{1-q_\star}}\right)}. \tag{30}$$

This result is presented as the *equilibrium* curve in Figure 2 *(Left)* of the main text.

## III. FRANZ-PARISI ENTROPY

In last Section we obtained some analytical proving that typical BCs of the stochastic perceptron can achieve essentially zero training error if $\beta$ and $q_*$ are large enough, and if the load $\alpha$ is below some critical capacity. This BCs are therefore solution of the binary perceptron problem. While typical (most numerous) solutions of the binary solutions problem are known to be isolated [17], we will show here that typical BCs belong to the dense solution region uncovered in Ref. [5]. Notice that, while for $q_* = 1$ the stochastic perceptron reduces to binary one, the limit $q_* \to 1$ of many observables will not be continuous due to this phenomena. Most noticeably, as shown in [5], the generalization error of solutions in the dense region is typically lower than the generalization error of isolated solutions.

We are interested in counting the number of solutions of the binary perceptron problem at fixed Hamming distance $d$ from a typical BC of the stochastic perceptron. For notation convenience we work at fixed overlap $p = \frac{1}{N}\sum_i W_i \,\mathrm{sign}(m_i)$, which can be linked to $d$ by the relation $p = 1 - 2d$. Following [12, 17] we define the Franz-Parisi entropy as

$$\mathcal{S}(p) = \lim_{N\to\infty} \frac{1}{N}\mathbb{E}_\mathcal{D}\left\langle \log\left[\sum_W \prod_{(x,y)\in\mathcal{D}} \Theta\left(y\sum_i W_i x_i\right) \times \delta\left(pN - \sum_i \mathrm{sign}\left(m_i\right) W_i\right)\right]\right\rangle, \tag{31}$$

where the expectation $\langle\bullet\rangle$ over $m$ is defined as usual according to Gibbs measure of the stochastic perceptron given by Eq. (1). The sum $\sum_W$ is over the binary configuration space $\{-1,+1\}^N$. The computation of $\mathcal{S}(p)$ is lengthy but straightforward, and can be done along the lines of Refs. [12, 17] using the replica method within the RS ansatz. Here we have the additional complication of some extra order parameters, due to the presence of the signin the constraint. We will present here just the final result. First, the order parameters $q_0, \hat{q}_0$ and $\hat{q}_1$ can be independently fixed solving the saddle point equations (15). $\mathcal{S}(p)$ is then given by

$$\mathcal{S}(p) = \underset{Q_0,\hat{Q}_0,s_0,s_1,\hat{s}_0,\hat{s}_1,\hat{p}}{\mathrm{extr}} \; -\frac{1}{2}\hat{Q}\left(1-Q\right) + \hat{s}_0 s_0 - \hat{s}_1 s_1 - \hat{p}p + G_S^{FP}(\hat{Q}_0,\hat{s}_0,\hat{s}_1,\hat{p}) + \alpha G_E^{FP}(Q_0,s_0,s_1), \tag{32}$$

where the entropic contribution is given by

$$G_S^{FP}(\hat{Q}_0,\hat{s}_0,\hat{s}_1,\hat{p}) = \int \mathcal{D}z \frac{\int_{-1}^{1} \mathrm{d}m \int \mathcal{D}\eta \; e^{\frac{1}{2}(\hat{q}_1 - \hat{q}_0)\tilde{W}^2 + \sqrt{\hat{q}_0}zm} A\left(m,\eta,z\right)}{\int_{-1}^{1} \mathrm{d}m \; e^{\frac{1}{2}(\hat{q}_1-\hat{q}_0)m^2 + \sqrt{\hat{q}_0}zm}}, \tag{33}$$

with

$$A\left(m,\eta,z\right)=\log 2\cosh\left(\left(\hat{s}_1-\hat{s}_0\right)m+\hat{p}\,\mathrm{sign}\left(m\right)+\sqrt{\frac{\hat{Q}_0\hat{q}_0-\hat{s}_0^2}{\hat{q}_0}}\eta+\frac{\hat{s}_0}{\sqrt{\hat{q}_0}}z\right),\tag{34}$$

and the energetic contribution by

$$G_E^{FP}\left(Q_0,s_0,s_1\right)=\int\mathcal{D}z_0\frac{\int\mathcal{D}\eta\,\mathcal{D}z_1\;H^{\beta}\left(-\frac{\sqrt{q_0}z_0+\sqrt{a}z_1+\frac{s_1-s_0}{\sqrt{b}}\eta}{\sqrt{1-q_\star}}\right)\log H\left(-\frac{\sqrt{b}\eta+\frac{s_0}{\sqrt{q_0}}z_0}{\sqrt{1-Q_0}}\right)}{\int\mathcal{D}z_1\;H^{\beta}\left(-\frac{\sqrt{q_0}z_0+\sqrt{q_\star-q_0}z_1}{\sqrt{1-q_\star}}\right)},\tag{35}$$

with

$$a=q_\star-q_0-\frac{\left(s_1-s_0\right)^2}{Q_0-s_0}\left(1-\frac{s_0\left(q_0-s_0\right)}{Q_0q_0-s_0^2}\right);\qquad b=\frac{Q_0q_0-s_0^2}{q_0}.\tag{36}$$

The extremization condition of Eq. (32) reads

$$\hat{Q}_0=-2\alpha\frac{\partial G_E^{FP}}{\partial Q_0};\;\hat{s}_0=-\alpha\frac{\partial G_E^{FP}}{\partial s_0};\;\hat{s}_1=\alpha\frac{\partial G_E^{FP}}{\partial s_1};\tag{37}$$

$$Q_0=1-2\frac{\partial G_S^{FP}}{\partial\hat{Q}_0};\;s_0=-\frac{\partial G_S^{FP}}{\partial\hat{s}_0};\;s_1=\frac{\partial G_S^{FP}}{\partial\hat{s}_1};\;0=\frac{\partial G_S^{FP}}{\partial\hat{p}}-p.\tag{38}$$

This system of coupled equations can be solved once again by iteration, with last equation being solved for $\hat{p}$ at each step with Newton method. The solution can then be plugged into Eq. (32), thus obtaining the final expression for the Franz-Parisi entropy. In Figure 2 *(Right)* of the main text we show the results for $\mathcal{S}(p)$ at $\alpha=0.55$, $\beta=20$ and different values of $q_\star$. Due to convergence problems in finding the fixed point of Eqs. (37,38), some of the curves could not be continued to large values of $d=(1-p)/2$. It is not clear if the problem is purely numerical and caused by the many integrals appearing in the equations and by the large value of $\beta$, or if it is an hint of a replica symmetry breaking transition. Nonetheless the region we are interested in exploring is that of low $d$, where the curve at $q_*=0.9$ reaches the origin, meaning that typical BCs are in the dense region of binary solution at this point. In the same figure we also compare the $\mathcal{S}(p)$ with two similar Franz-Parisi entropies, which we denote here by $\mathcal{S}_{\mathrm{bin}}(p)$ and $\mathcal{S}_{\mathrm{sph}}(p)$, for the binary and the spherical perceptron respectively. These two entropies are defined as in Eq. (31), the only difference being that the expectation $\langle\bullet\rangle$ over $m$ is according to

$$Z=\int d\nu(m)\prod_\mu\Theta\Big(y^\mu\sum_i m_i x_i^\mu\Big),\tag{39}$$

with $d\nu(m)=\prod_i(\delta(m_i-1)+\delta(m_i-1))dm_i$ in the binary case and $d\nu(m)=\prod_i dm_i\;\delta\big(\sum_i m_i^2-N\big)$ in the spherical one.

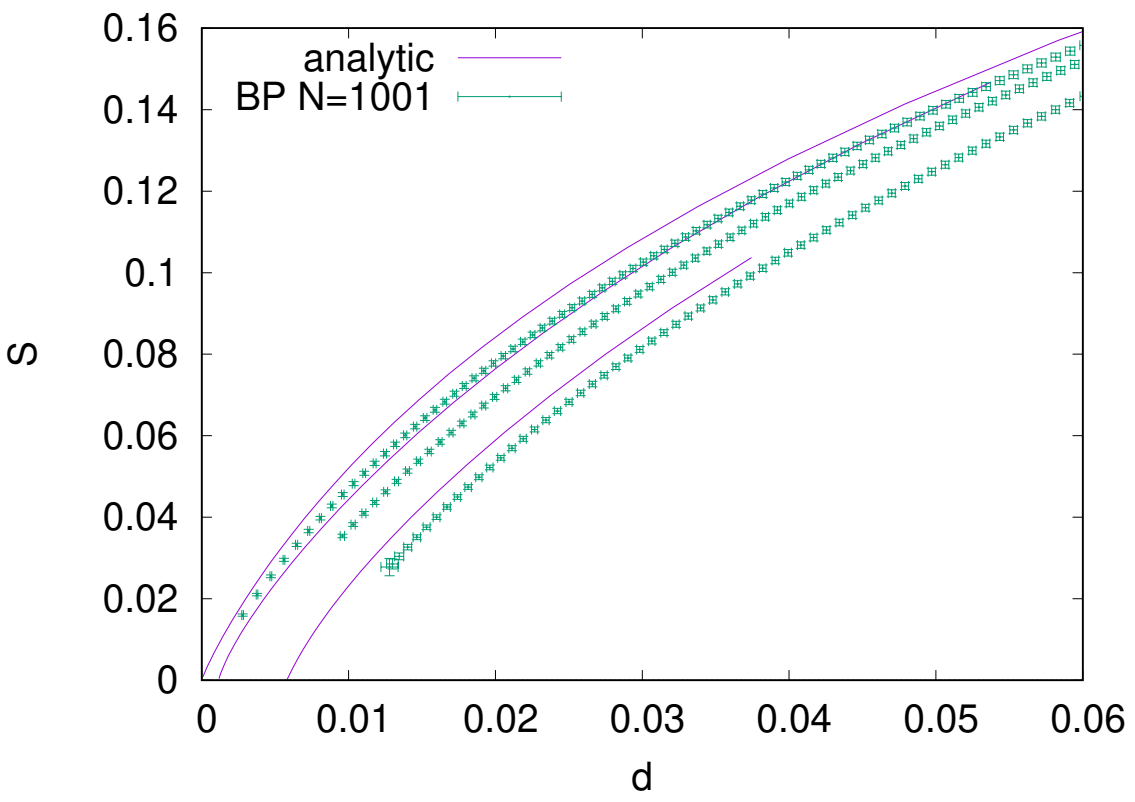

Figure 2. Franz-Parisi entropies for $\alpha = 0.55$ and $q_\star = 0.7, 0.8, 0.9$ (from top to bottom). (Purple) Average case Franz-Parisi entropy $\mathcal{S}(d)$ as given by Eq. Eq. (32) for $\beta = 20$. (Green) Single sample Franz-Parisi entropies computed with Belief Propagation, averaged over 100 samples.

We also derived and implemented a single sample version of the Franz-Parisi calculation, performed as follows. For a given realization of $\mathcal{D}$ we establish a slow schedule of $q_\star$ values and perform a GD on $m^t$ where after each update we rescale the squared norm of $m^t$ to $q_\star N$ until convergence, before moving to the next value of $q_\star$. At any point of the schedule, the configuration $m^t$ is binarized and given and the constrained entropy of the binary perceptron is computed using the Bethe approximation given by the Belief Propagation algorithm (see Ref. [6] for a detailed exposition). The result of the simulations are presented in Fig. 2. The slight deviation of the BP results from the analytic curves could be explained by several causes: 1) finite size effects; 2) the analytic prediction is for non-zero (although low) temperature; 3) the reference configuration $m^t$ is selected through the GD dynamics, while in the analytic computation $m$ is sampled according to the thermal measure defined by partition function of Eq. (1).

## IV. BINARY CONTROL VARIABLES

In order to make a straightforward connection with the large deviation analyses proposed in Ref. [5], we have also considered the case in which the control variables $m_i$ are discretized as well: $m_i = \sqrt{q_\star}\sigma_i$, with $\sigma_i \in \{-1, 1\}$. In this case the log-likelihood of the stochastic perceptron model reads:

$$\mathcal{L}(\sigma) = \sum_{\mu=1}^{M} \log H\left(-\sqrt{\frac{q_\star}{1-q_\star}}\frac{y^\mu \sum_i \sigma_i x_i^\mu}{\sqrt{\sum_i (x_i^\mu)^2}}\right). \tag{40}$$

The statistical physics analysis proposed in the main text can be easily adapted to this case. Fig. (3) shows the average energy $E$, associated to a typical configuration $\sigma$, as a function of $q_\star$. The analytic results are found to be in reasonable agreement with the estimation of the training error obtained through a Markov Chain Monte Carlo on the system with Hamiltonian given by $-\mathcal{L}(\sigma)$, with inverse temperature $\beta = 15$.

Moreover, instead of assuming $\sigma$ to be the parameters controlling $Q_m(W)$, from which the stochastic binary synapses are sampled, it is instructive to take a different perspective: consider a model where the synapses are binary and deterministic, and where we introduce a dropout mask [21], randomly setting to zero a fraction $p$ of the inputs. In this scenario, we can write the log-likelihood of obtaining a correct classification over the independent realizations of the dropout mask for each datapoint. For large $N$ the resulting log-likelihood is exactly that given by Eq. (40), once we set $q_\star = 1 - p$. Moreover, in the case of a single layer network we are considering, the dropout mask on the inputs could be equivalently applied to the synapses, as in the drop-connect scheme [22]. We can thus see a clear analogy between the dropout/dropconnect schemes and the learning problem analyzed throughout this paper, even though in standard machine learning practice the synaptic weights $\sigma_i$ are not constrained to binary values.

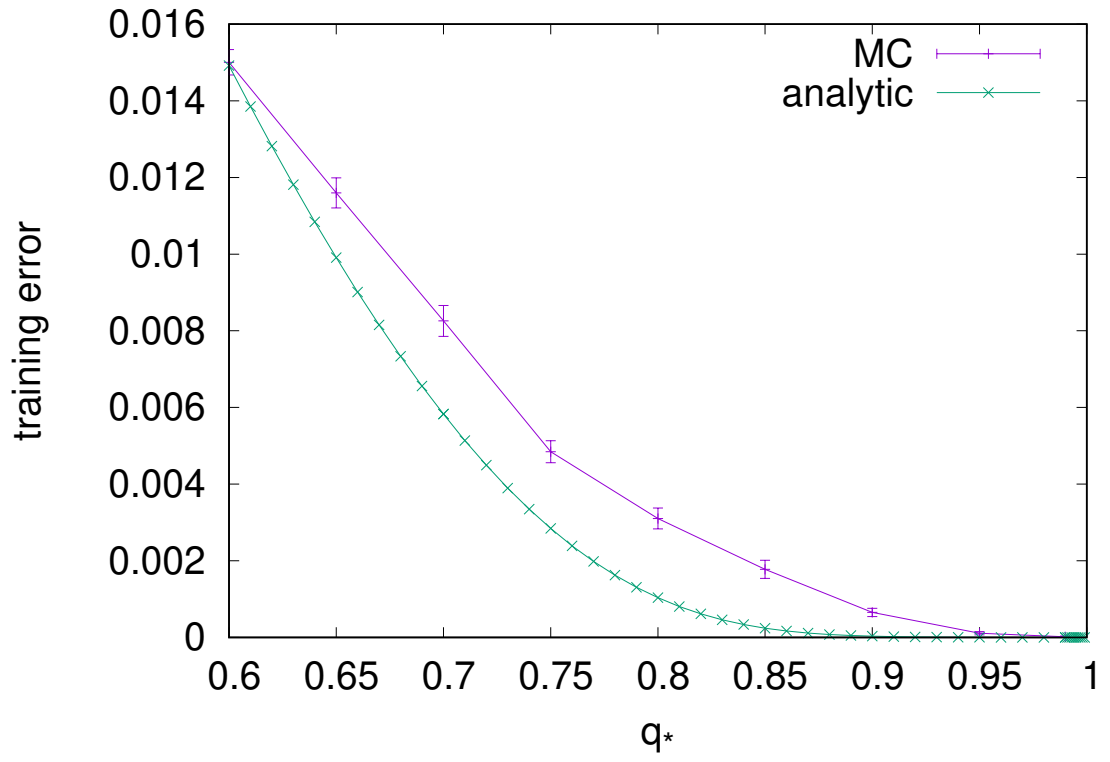

Figure 3. Stochastic perceptron with binary control. Energy of the clipped center versus $q_\star$. Red curve, MC simulation at $N = 1001$, averaged over 100 samples. Green curve, analytic results determined through the replica approach. Storage load $\alpha = 0.55$, inverse temperature $\beta = 15$.

## V. STOCHASTIC DEEP NETWORKS

The stochastic framework can be extended to train deep networks models with binary synapses and binary activations using standard deep learning techniques, once some approximations to the log-likelihood estimation are taken into account. Since this extension is beyond the scope of the present paper, here we only sketch the training algorithm and give some preliminary results on its performance, reserving a detailed overview and extensive testing to a future publication [3].

Consider a multi-layer perceptron with $L$ hidden neuron layers, with synaptic weights $W_{ij}^{\ell}$, $\ell = 0, \dots, L$, and sign activations:

$$\tau_j^{\ell+1} = \text{sign}\left(\sum_j W_{ij}^{\ell} \tau_j^{\ell} + b_i^{\ell}\right), \quad \ell = 0, \dots, L, \tag{41}$$

where $\tau^0 = x$ is the input of the network, and $b^\ell$ are continuous biases to be optimized. In our stochastic framework the weights $W_{ij}^{\ell}$ are independent binary ($\pm 1$) random variables with means $m_{ij}^{\ell}$ to be optimized. For a fixed activation trajectory $(\tau^\ell)_\ell$ and wide layers, expectations with respect to $W$ can be taken. Also, adapting the scheme of Ref. [15], the probabilistic iteration of the neuron activation distribution $P_\ell(\tau^\ell)$ across the layers can be performed within a factorized approximation, in terms of the neuron activation's means $a_i^\ell$:

$$a_i^{\ell+1} = 2H\left(-\frac{\sum_j m_{ij}^\ell a_j^\ell + b_i^\ell}{\sum_j 1-(m_{ij}^\ell)^2(a_j^\ell)^2}\right) - 1$$

Finally an approximated softmax layer can be defined on the last layer output $a^{L+1}$, and consequently a cross-entropy loss function can be used in the training. We experimented this approach on the MNIST dataset, where we trained networks a 3 hidden layers of width 801. We approximately minimized the loss function using Adam optimizer with standard parameters and learning rate $\eta = 10^{-2}$. We used in our simulations the Julia deep learning library Knet [23], providing automatic differentiation, backpropagation and GPU acceleration. At the end of the training the resulting binarized configuration, $\hat{W}_{ij}^\ell = \text{sign}(m_{ij}^\ell)$, achieved $\sim 1.7\%$ test error. Our implementation of the current state of the art algorithm [10] on the same network, using batch normalization and with learning rate $\eta = 10^{-3}$, achieves approximately the same result. For the sake of comparison, we note that a standard neural network with the same structure but with ReLU activations and continuous weights we obtain $\sim 1.4\%$ test error. Given the heavy constraints on the weights, the discontinuity of the sign activation and the peculiarity of the training procedures, it is quite astonishing to observe only a slight degradation in the performance of binary networks when compared to their continuous counterparts. Further improvements of our results, up to $\sim 1.3\%$ test error, can be obtained applying dropout on the input and the hidden layers.

## VI. A WEIGHTED PERCEPTRON RULE

In the main text we introduced a stochastic perceptron model where the stochastic synapses could be integrated out thanks to the central limit theorem. Therefore we could express the log-likelihood of the model $\mathcal{L}(m)$ as an easy to compute function of the parameters $m$ governing the distribution $Q_m(W) = \prod_{i=1}^N \left[\frac{1+m_i}{2}\delta_{W_i,+1} + \frac{1-m_i}{2}\delta_{W_i,-1}\right]$. We used deterministic gradient ascent as a procedure to optimize $m$. At convergence, the binarized configuration $W_i = \text{sign}(m_i)$ is proposed as an approximate solution of the binary problem. This learning rule (i.e. Eq. (7) in the main text) can be rewritten as

$$m_i' = \text{clip}\left(m_i + \eta\frac{1}{M}\sum_{\mu=1}^M K\left(-\frac{y^\mu \overline{h}^\mu}{\overline{\sigma}^\mu}\right)\left(\frac{y^\mu x_i^\mu}{\overline{\sigma}^\mu} + \frac{(x_i^\mu)^2\overline{h}^\mu}{(\overline{\sigma}^\mu)^3}\right)\right), \tag{42}$$

where we defined $\overline{h}^\mu = \sum_i m_i x_i^\mu$, $\overline{\sigma}^\mu = \sqrt{\sum_i (1-m_i^2)(x_i^\mu)^2}$ and $K(x) = \partial_x \log H(x)$.

We now proceed to modify the learning rule to test its adaptability to biological scenarios. As a premise, we note that the emergence of a discretized set of synaptic strengths, as encoded by our model, is an experimentally observed property of many neural systems [8, 19]. Inspection of (42) shows an Hebbian structure, where the synaptic strength is reinforced on the base of presynaptic and postsynaptic activity, with a modulating factor $K(-y^\mu\overline{h}^\mu/\overline{\sigma}^\mu)$ that can be interpreted as a reward signal [18].

The sum over the examples in the training set can be changed with the random extraction of a single index $\mu$. In this way the algorithm can be naturally extended to an online learning scenario. We revert to the original stochastic variables, sampling $W_i \sim Q_{m_i}$ $\forall i$ and we replace the average pre-activation value $\overline{h}^\mu$ with its realization $h^\mu = \sum_i W_i x_i^\mu$. Further butchering of (42) is obtained crudely replacing $\overline{\sigma}^\mu$ by the constant $\sigma = \sqrt{0.5N}$. The final stochastic rule reads

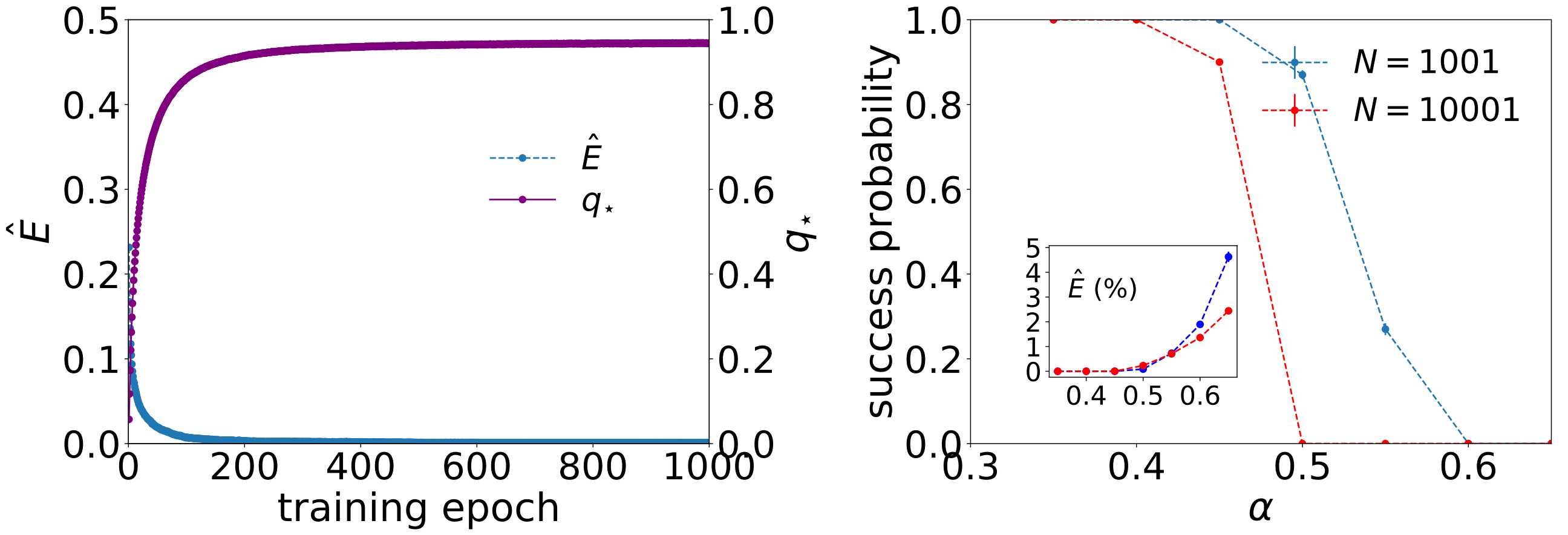

Figure 4. Performance of learning rule Eq. (43). Results on system size $N = 1001$ are averaged over 100 samples, learning rate $\eta = 10^{-2}\sigma$. Experiments at $N = 10001$ are averaged over 10 samples, learning rate $\eta = 10^{-3}\sigma$. (*Left*) The training error and the squared norm against the number of training epochs for $\alpha = 0.55$ and $N = 1001$. (*Right*) Success probability within 2000 epochs in the classification task as a function of the load $\alpha = M/N$. In the inset we show the average training error (in percentage) at the end of GD.

$$m_i' = \text{clip}\left(m_i + \eta K\left(-\frac{y^\mu h^\mu}{\sigma}\right)\left(\frac{y^\mu x_i^\mu}{\sigma} + \frac{(x_i^\mu)^2 h^\mu}{\sigma^3}\right)\right). \tag{43}$$

We measure the performance of rule (43), on randomly genererated training sets with uniform i.i.d. $x_i^\mu = \pm 1$ and $y^\mu = \pm 1$. We present the results in Fig. 4.

We observe a degradation in performance with respect to rule (42) and longer convergence times. Nonetheless, the algorithm is still able to efficiently classify an extensive number of examples (for the considered system sizes) up to a load $\alpha = M/N \sim 0.45$. As with gradient descent, above the algorithmic threshold we observe a graceful increase of the training error of the binarized configurations returned by the algorithm.

Learning rule (43) could be utterly simplified if we discarded the last term on the right hand side and we replaced $K(x)$ with the Heaviside theta function $\Theta(x)$, which takes value 0 for negative arguments and 1 otherwise. The new rule would read

$$m_i' = \text{clip}\left(m_i + \eta\Theta\left(-y^\mu h^\mu\right) y^\mu x_i^\mu\right). \tag{44}$$

We first observe that, if we choose $h^\mu = \sum_i \text{sign}(m_i) x_i^\mu$, we obtain a variant of the clipped perceptron (CP) algorithm, analyzed in Refs. [1, 4]. The performances of this rule were shown to degrade rapidly with $N$. For example, rule (44) with deterministic $h^\mu$ fails to find solution within 2000 epochs at $N = 2001$ and $\alpha = 0.3$. Instead, we find that if we consider $h^\mu = \sum_i W_i x_i^\mu$, with $W_i$ sampled according to $m_i$, we obtain a stochastic version of the rule able to succeed in the same setting. We note also that the ordinary perceptron rule, $m_i' = \text{clip}\left(m_i + \eta\Theta(-y^\mu \overline{h}^\mu) y^\mu x_i^\mu\right)$, is not able to provide binary solutions, even at very low $\alpha$.

Although a proper assessment of the scaling behaviour of these algorithms with $N$ is beyond the scope of this work, we report that rule (43) consistently outperforms both variants of rule (44). Moreover its performance can be further improved using the actual value $\overline{\sigma}^\mu$ instead of $\sigma$. In the next section we show that stochastic variant of (44), which is a novel contribution of this paper and it is tied to the stochastic framework we have investigated, typically lands in a flatter region compared to the deterministic version.

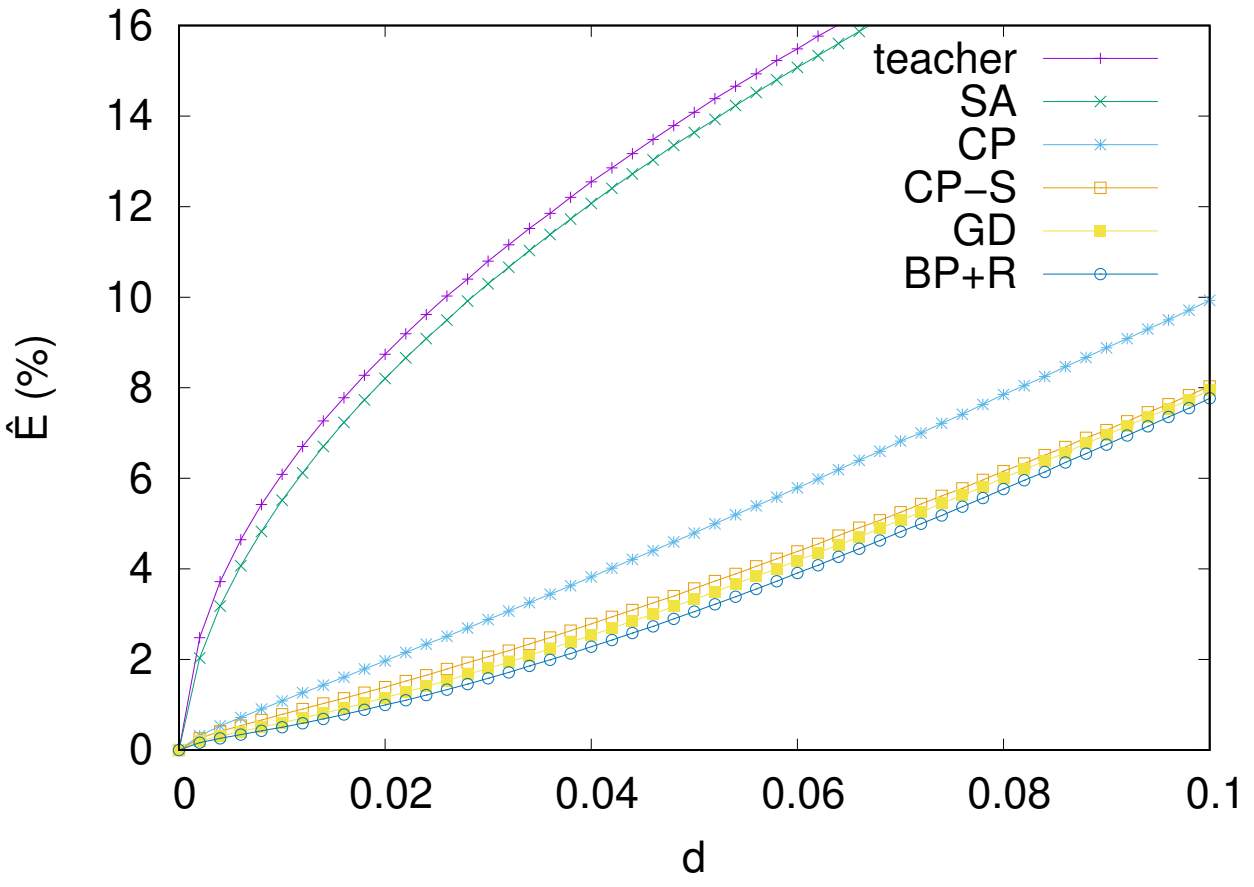

Figure 5. The energy of Eq. 45 as a function of the Hamming distance $dN$ from the teacher and from solutions found by different algorithms. $N = 1001$ and $\alpha = 0.4$ in all simulations. Curves are averaged over 40 samples.

## VII. AVERAGE ENERGY AROUND ALGORITHMIC SOLUTIONS

In order to characterize the energy landscape around a reference configuration $W$, a standard tool is the constrained entropy $\mathcal{S}(W,d)$ (also called local entropy), counting the number of solutions (i.e. zero energy configurations) at distance $d$ from $W$. The average properties of $\mathcal{S}(W,d)$ have been analyzed in the main text. If for any small $d > 0$ we have $\mathcal{S}(W,d) > 0$, we say that the configuration $W$ belongs to a dense cluster. Otherwise, we say that $W$ is isolated [5, 17].

Along with $\mathcal{S}(W,d)$, we can consider a simpler observable that can help to build a picture of an heterogeneous energy landscape, made of wide and sharp basins. Following Ref. [7], we thus define the average misclassification error made by configurations at distance $d$ from $W$, as

$$\hat{E}(W,d) = \mathbb{E}_{W'|W,d}\ \frac{1}{M}\sum_{\mu=1}^{M}\Theta\left(-y^\mu\sum_i W_i' x_i^\mu\right), \tag{45}$$

where the expectation is defined by

$$\mathbb{E}_{W'|W,d}\bullet = \frac{\sum_{W'}\bullet\times\delta\left(N\left(1-2d\right)-\sum_i W_i W_i'\right)}{\sum_{W'}\delta\left(N\left(1-2d\right)-\sum_i W_i W_i'\right)} \tag{46}$$

Notice that configurations $W'$ partecipating to $\mathbb{E}_{W'|W,d}$ are not required to be solutions of the problem: we can easily sample them by choosing $dN$ spins at random to be flipped in $W$. In our tests the expectation is approximated by $10^3$ Monte Carlo samples.

| Teacher | SA | CP | CP-S | GD | BP+R |
|---|---|---|---|---|---|
| 1 | 0.578(3) | 0.628(3) | 0.644(3) | 0.642(3) | 0.657(2) |

Table I. Generalization accuracy in the teacher-student scenario. $N = 1001$, $\alpha = 0.4$, averages over 40 samples.

We explored the behavior of $\hat{E}(W, d)$ as a function of $d$ for different solutions $W$ of the problem, obtained from different algorithms. We compare: the Gradient Descent (GD) algorithm investigated in the main text (Eq. 7); the two variants of rule 44 (from previous section), the one with deterministic $h^\mu = \sum_i \mathrm{sign}(m_i) x_i^\mu$ (CP) and the one with stochastic $h^\mu$ sampled according to $m$ (CP-S); the Belief Propagation algorithm with reinforcement heuristic (BP+R) of Ref. [9]; Simulated Annealing (SA) on the Hamiltonian $\sum_\mu \Theta_1(-y^\mu \sum_i W_i' x_i^\mu / \sqrt{N})$, where $\Theta_1(x) = x\,\Theta(x)$.

In order to compare the properties of the algorithmic solutions and the typical isolated solutions, it is useful to consider the so-called teacher-student scenario [11]. As before, we generate uniformly i.i.d. $x_i^\mu = \pm 1$, but we assign the labels according to a teacher configuration $W^T$(we can choose $W_i^T = 1\,\forall i$ without loss of generality). In this scenario, the teacher has the same statistical properties of the typical solutions of the training set, therefore it is an isolated solution itself [5, 17].

Results are presented in Fig. 5. For the isolated teacher we see a rapid increase of the average energy around the origin. The same happens for solutions discovered by SA, which we can classify as isolated as well. Also, SA was the slowest algorithm to reach a solution (unsurprisingly, since it is known to scale badly with the system size [2, 16, 17]).

Solutions from CP-S, GD and BP+R instead are surrounded by a much flatter average landscape and, remarkably, they all give similar results. These three algorithms are implicitly or explicitly devised to reach robust basins: GD and CP-S are derived within our stochastic framework, while the reinforcement term in BP+R has been shown in [2] to be linked to local entropy maximization. Solutions from CP algorithm, while not in basins as sharp as the ones found by SA, do not achieve the same robustness as ones from those three algorithms.

We give some additional details on the simulation's protocol. The setting chosen, $N = 1001$ and $\alpha = 0.4$ (a low load regime in the teacher-student scenario), is such that each algorithm finds a solution on each instance within small computational time (a few minutes). As soon as a solution $W$ is discovered the algorithm is arrested. Results could be slightly different for some algorithms if they were allowed to venture further within the basin of solutions. For CP and CP-S we set $\eta = 2 * 10^{-3}$, while $\eta = 0.1$ for GD. The reinforcement parameter in BP+R is updated according to $1 - r^{t+1} = (1 - r^t)(1 - 10^{-3})$ while the inverse temperature schedule in SA is $\beta^{t+1} = \beta^t (1 + 5 * 10^{-3})$ .

In Table I we report the probability of correctly classifying a new example generated by the teacher, usually called generalization accuracy, for each of the algorithms investigated. We note a clear correlation between the flatness of the basin as seen in Fig. 5 and the ability to classify correctly new examples, with SA having the worst performances and BP+R the best ones.